# PRECISION TESTS OF QUANTUM CHROMODYNAMICS AND THE STANDARD MODEL *

STANLEY J. BRODSKY

*Stanford Linear Accelerator Center*
*Stanford University, Stanford, California 94309*

HUNG JUNG LU

*Department of Physics, University of Arizona*
*Tucson, Arizona 20742*

We discuss three topics relevant to testing the Standard Model to high precision: commensurate scale relations, which relate observables to each other in pertubation theory without renormalization scale or scheme ambiguity, the relationship of compositeness to anomalous moments, and new methods for measuring the anomalous magnetic and quadrupole moments of the $W$ and $Z$.

## 1. Introduction

One of the obstacles to testing the Standard Model to high precision is the fact that perturbative predictions depend on the choice of rernormalization scale and scheme. The sitution is further complicated by the fact that computations in different sectors of the Standard Model are carried out using different renormalization schemes. For example, in quantum electrodynamics, higher order radiative corrections are computed in the traditional "on-shell" scheme using Pauli-Villars regularization. The QED coupling $\alpha_{\rm QED}$ is defined from the Coulomb scattering of heavy test charges at zero momentum transfer. The scale $k^2$ in the running QED coupling $\alpha_{\rm QED}(k^2)$ is then set by the virtuality of the photon propagator in order to sum all vacuum polarization corrections. However, in the non-Abelian sectors of the Standard Model, higher order computations are usually carried out using the $\overline{\rm MS}$ dimensional regularization scheme. The renormalization scale $\mu$ that appears in perturbative expansions in the QCD coupling $\alpha_{\overline{\rm MS}}(\mu^2)$ is usually treated as an arbitrary parameter. These ambiguities and disparities in choices of scales and schemes lead to uncertainties in establishing the accuracy and range of validity of perturbative QCD predictions and in testing the hypothesis of grand unification.

In this talk, we shall discuss a set of new high precision perturbative predictions for the Standard Model which have no scale or scheme ambiguities. These predictions, called "Commensurate Scale Relations"[1], are valid for any renormalizable quantum field theory, and thus may provide a uniform perturbative analysis of the electroweak and strong sectors of the Standard Model. Commensurate scale relations relate observables to observables, and thus must be independent of theoretical conventions, such as choice of intermediate renormalization scheme. The scales of the effective charges that appear in commensurate scale relations are fixed by the requirement that the couplings sum all of the effects of the non-zero $\beta$ function, as in the BLM method [2]. The coefficients in the perturbative expansions in the commensurate scale relations are thus identical to those of a corresponding conformally-invariant theory with $\beta = 0$. The scales that appear in commensurate scale relations are physical since they reflect the mean virtuality of the gluons in the underlying hard subprocess [3]. As emphasized by Mueller [4] at this conference, commensurate scale relations isolate the effect of infrared renormalons associated with the non-zero $\beta$ function. The usual factorial growth of the coefficients in perturbation theory due to quark and gluon vacuum polarization insertions is eliminated since such effects are resummed into the running couplings. The perturbative series is thus much more convergent. In the next section we discuss an elegant example: a surprisingly simple connection between the radiative corrections to the Bjorken sum rule and the radiative corrections to the $e^+e^-$ annihilation cross section. The coefficients that appear in the perturbative expansion are a simple geometric series. This relation generalizes Crewther's relation to non-conformal QCD.

Commensurate scale relations can also be applied in grand unified theories to make scale and scheme invariant predictions which relate physical observables in different sectors of the theory. In addition, the commensurate scale relation between $\alpha_V$, as defined from the heavy quark potential, and $\alpha_{\overline{MS}}$ provides an analytic extension of the $\overline{MS}$ scheme in which flavor thresholds are taken into account the proper scale automatically. The heavy quark coupling $\alpha_V$ has been recently been determined to high precision from lattice gauge theory [5] by using an improved perturbation theory closely related to the BLM method.

In the Standard Model, it is assumed that the lepton, quark, and vector bosons are all elementary. In the second part of this talk, we discuss the ways in which a composite spin-$\frac{1}{2}$ or spin-1 system can mimic the quantum of an elementary field, provided its size $R$, defined from the slope of form factors, is small compared to its Compton scale $1/M$. In particular, we shall use a light-cone description of relativistic bound states to show that the anomalous moment of a composite system vanishes in the point-like $MR \to 0$ limit [6]. The light-cone Fock state method also provides an important relationship between the axial coupling and magnetic moment of a composite system.

One of the remarkable consequences of the canonical couplings of the Standard Model is a superconvergent sum rule for polarized photoabsorption cross sections at the tree level [7,8]. This classical sum rule in turns imply the reversal of sign of the polarization asymmetry at a specific energy for processes such as $\gamma e^- \to W^- \nu_e$ [9]. The implications of these predictions for high precision tests of the Standard Model and limits on compositeness are discussed in Section 4.

## 2. Commensurate Scale Relations and The Generalized Crewther Relation in Quantum Chromodynamics

In 1972 Crewther [10] derived a remarkable consequence of the operator product expansion for conformally-invariant gauge theory. Crewther's relation has the form

$$3S = KR' \tag{1}$$

where $S$ is the value of the anomaly controlling $\pi^0 \to \gamma\gamma$ decay, $K$ is the value of the Bjorken sum rule in polarized deep inelastic scattering, and $R'$ is the isovector part of the annihilation cross section ratio $\sigma(e^+e^- \to \text{hadrons})/\sigma(e^+e^- \to \mu^+\mu^-)$. Since $S$ is unaffected by QCD radiative corrections [11], Crewther's relation requires that the QCD radiative corrections to $R_{e^+e^-}$ exactly cancel the radiative corrections to the Bjorken sum rule order by order in perturbation theory.

However, Crewther's relation is only valid in the case of conformally-invariant gauge theory, *i.e.* when the coupling $\alpha_s$ is scale invariant. This is apparent since the radiative corrections to the Bjorken sum rule and the annihilation ratio are in general functions of different physical scales. Thus Crewther's relation cannot be tested directly in QCD unless the effects of the nonzero $\beta$ function for the QCD running coupling are accounted for, and the energy scale $\sqrt{s}$ in the annihilation cross section is related to the momentum transfer $Q$ in the deep inelastic sum rules. Recently Broadhurst and Kataev [12] have explicitly calculated the radiative corrections to the Crewther relation and have demonstrated explicitly that the corrections are proportional to the QCD $\beta$ function.

A helpful tool for relating physical quantitities is the effective charge. Any perturbatively calculable physical quantity can be used to define an effective charge [13,14,15] by incorporating the entire radiative correction into its definition. An important result is that all effective charges $\alpha_A(Q)$ satisfy the Gell-Mann-Low renormalization group equation with the same $\beta_0$ and $\beta_1$; different schemes or effective charges only differ through the third and higher coefficients of the $\beta$ function. Thus, any effective charge can be used as a reference running coupling constant in QCD to define the renormalization procedure. More generally, each effective charge or renormalization scheme, including $\overline{\text{MS}}$, is a special case of the universal coupling function $\alpha(Q, \beta_n)$ [16]. Peterman and Stückelberg have shown [17] that all effective charges are related to each other through a set of evolution equations in the scheme parameters $\beta_n$.

For example, consider the entire radiative corrections to the annihilation cross section expressed as the "effective charge" $\alpha_R(Q)$ where $Q = \sqrt{s}$:

$$R(Q) \equiv 3 \sum_f Q_f^2 \left[1 + \frac{\alpha_R(Q)}{\pi}\right]. \tag{2}$$

Similarly, we can define the entire radiative correction to the Bjorken sum rule as the effective charge $\alpha_{g_1}(Q)$ where $Q$ is the lepton momentum transfer:

$$\int_0^1 dx \left[ g_1^{ep}(x, Q^2) - g_1^{en}(x, Q^2) \right] \equiv \frac{1}{3} \left| \frac{g_A}{g_V} \right| \left[ 1 - \frac{\alpha_{g_1}(Q)}{\pi} \right]. \qquad (3)$$

We now use the known expressions to three loops [18,19,20] in $\overline{\text{MS}}$ scheme and choose the leading-order and next-to-leading scales $Q^*$ and $Q^{**}$ to re-sum all quark and gluon vacuum polarization corrections into the running couplings. The values of these scales are the physical values of the energies or momentum transfers which ensure that the radiative corrections to each observable passes through the heavy quark thresholds at their respective commensurate physical scales. The final result is remarkably simple:

$$\frac{\alpha_{g_1}(Q)}{\pi} = \frac{\alpha_R(Q^*)}{\pi} - \left( \frac{\alpha_R(Q^{**})}{\pi} \right)^2 + \left( \frac{\alpha_R(Q^{***})}{\pi} \right)^3 + \cdots. \qquad (4)$$

The coefficients in the series (aside for a factor of $C_F$, which can be absorbed in the definition of $\alpha_s$) are actually independent of color and are the same in Abelian, non-Abelian, and conformal gauge theory. The non-Abelian structure of the theory is reflected in the scales $Q^*$ and $Q^{**}$. Note that the $\overline{\text{MS}}$ renormalization scheme is used here for calculational convenience; it serves simply as an intermediary between observables. This is equivalent to the group property defined by Peterman and Stückelberg [17] which ensures that predictions in PQCD are independent of the choice of an intermediate renormalization scheme. (The renormalization group method was developed by Gell-Mann and Low [21] and by Bogoliubov and Shirkov [22].)

The connection between the effective charges of observables such as Eq. (4) is referred to as a "commensurate scale relation" (CSR). A fundamental test of QCD will be to verify empirically that the observables track in both normalization and shape as given by the CSR. The commensurate scale relations thus provide fundamental tests of QCD which can be made increasingly precise and independent of the choice of renormalization scheme or other theoretical convention. More generally, the CSR between sets of physical observables automatically satisfy the transitivity and symmetry properties [23] of the scale transformations of the renormalization "group" as originally defined by Peterman and Stückelberg [17]. The predicted relation between observables must be independent of the order one makes substitutions; *i.e.* the algebraic path one takes to relate the observables.

The relation between scales in the CSR is consistent with the BLM scale-fixing procedure[2] in which the scale is chosen such that all terms arising from the QCD $\beta$−function are resummed into the coupling. Note that this also implies that the coefficients in the perturbation CSR expansions are independent of the number of quark flavors $f$ renormalizing the gluon propagators. This prescription ensures that, as in quantum electrodynamics, vacuum polarization contributions due to fermion pairs are all incorporated into the coupling $\alpha(\mu)$ rather than the coefficients. The coefficients in the perturbative expansion using BLM scale fixing are the same as those of the corresponding conformally invariant theory with $\beta = 0$. In practice, the conformal limit is defined by $\beta_0, \beta_1 \to 0$, and can be reached, for instance, by adding enough spin-half and scalar quarks as in $N = 4$ supersymmetric QCD. Since all the running coupling effects have been absorbed into the renormalization scales, the BLM scale-setting method correctly reproduces the perturbation theory coefficients of the conformally invariant theory in the $\beta \to 0$ limit.

Let us now discuss in more detail the derivation of eqn. (4). The perturbative series of $\alpha_{g_1}(Q)/\pi$ using dimensional regularization and the $\overline{\rm MS}$ scheme with the renormalization scale fixed at $\mu = Q$ has been computed [18] through three loops in perturbation theory. The effective charge for the annihilation cross section has also been computed [19,20] to the same order in the $\overline{\rm MS}$ scheme with the renormalization scale fixed at $\mu = Q = \sqrt{s}$. The two effective charges can be related to each other by eliminating $\alpha_{\overline{\rm MS}}$. The scales $Q^*$ and $Q^{**}$ are set by resumming all dependence on $\beta = 0$ and $\beta_1$ into the effective charge. The application of the NLO BLM formulas then leads to

$$\frac{\alpha_{g_1}(Q)}{\pi} = \frac{\alpha_R(Q^*)}{\pi} - \frac{3}{4}C_F\left(\frac{\alpha_R(Q^{**})}{\pi}\right)^2$$

$$+ \left[\frac{9}{16}C_F^2 - \left(\frac{11}{144} - \frac{1}{6}\zeta_3\right)\frac{d^{abc}d^{abc}}{C_F N}\frac{\left(\sum_f Q_f\right)^2}{\sum_f Q_f^2}\right]\left(\frac{\alpha_R(Q^{***})}{\pi}\right)^3, \quad (5)$$

$$Q^* = Q\exp\left[\frac{7}{4} - 2\zeta_3 + \left(\frac{11}{96} + \frac{7}{3}\zeta_3 - 2\zeta_3^2 - \frac{\pi^2}{24}\right)\left(\frac{11}{3}C_A - \frac{2}{3}f\right)\frac{\alpha_R(Q)}{\pi}\right], \quad (6)$$

$$Q^{**} = Q\exp\left[\frac{523}{216} + \frac{28}{9}\zeta_3 - \frac{20}{3}\zeta_5 + \left(-\frac{13}{54} + \frac{2}{9}\zeta_3\right)\frac{C_A}{C_F}\right]. \quad (7)$$

In practice, the scale $Q^{***}$ in the above expression can be chosen to be $Q^{**}$. Notice that aside from the light-by-light contributions, all the $\zeta_3, \zeta_5$ and $\pi^2$ dependencies have been absorbed into the renormalization scales $Q^*$ and $Q^{**}$. Understandably, the $\pi^2$ term should be absorbed into renormalization scale since it comes from the analytical continuation of $R(Q)$ to the Euclidean region.

For the three flavor case, where we can neglect the light-by-light contribution, the series remarkably simplifies to the CSR of Eq. (4). The form suggests that for the general $SU(N)$ group the natural expansion parameter is $\hat{\alpha} = (3C_F/4\pi)\alpha$. The use of $\hat{\alpha}$ also makes it explicit that the same formula is valid for QCD and QED. That is, in the limit $N_C \to 0$ the perturbative coefficients in QCD coincide with the perturbative coefficients of an Abelian analog of QCD.

In Fig. 1 we plot the scales $Q^*$ and $Q^{**}$ as function of $Q$ for in the range $0 \le Q \le 6$. We can see that the scales $Q^*$ and $Q^{**}$ are of the same order as $Q$ but roughly a factor $1/2$ to $1/3$ smaller.

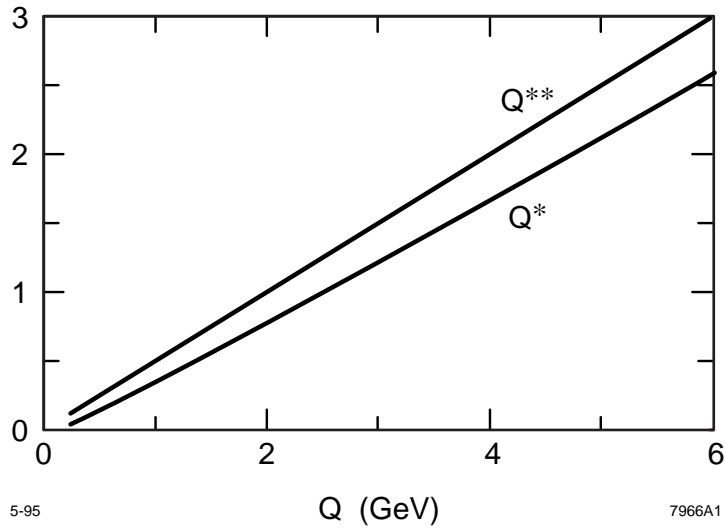

Figure 1: The commensurate scales $Q^*$ and $Q^{**}$ for the case of Bjorken sum rule expressed in terms of $\alpha_R(Q)$.

In Fig. 2 we plot the prediction for the value of the Bjorken sum rule using as input the values of $\alpha_R(Q)$ as given by Mattingly and Stevenson [24]. We use $Q^{***} = Q^{**}$ here. Notice that the prediction has a very smooth and flat behavior, even at $Q^2 \sim 2$ GeV$^2$ since the effective charge $\alpha_R(Q)$ as obtained by Mattingly and Stevenson incorporates the "freezing" of the strong coupling constant.

Broadhurst and Kataev have recently observed a number of interesting relations between $\alpha_R(Q)$ and $\alpha_{g_1}(Q)$ (the "Seven Wonders") [12]. In particular, they have shown the factorization of the beta function in the correction to Crewther's relation thus establishing a non-trivial connection between the total $e^+e^-$ annihilation cross section and the polarized Bjorken sum rule. The simple form of Eq. (4) also points to the existence of a "secret symmetry" between $\alpha_R(Q)$ and $\alpha_{g_1}(Q)$ which is revealed after the application of the NLO BLM scale setting procedure. In fact, as pointed out by Kataev and Broadhurst [12], in the conformally invariant limit, i.e., for vanishing beta functions, Crewther's relation becomes

$$(1 + \widehat{\alpha}_R^{\text{eff}})(1 - \widehat{\alpha}_{g_1}^{\text{eff}}) = 1. \qquad (8)$$

Thus Eq. (4) can be regarded as the extension of the Crewther relation to non-conformally invariant gauge theory.

The commensurate scale relation between $\alpha_{g_1}$ and $\alpha_R$ given by Eq. (4) implies that the radiative corrections to the annihilation cross section and the Bjorken (or Gross-Llewellyn Smith) sum rule cancel at their commensurate scales. The relations

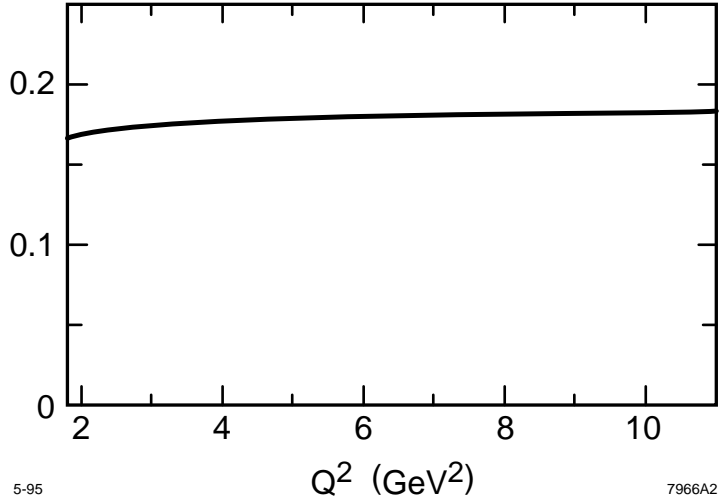

Figure 2: Prediction of the Bjorken sum rule from $R_{e^+e^-}$ according to the commensurate scale relation and using Mattingly and Stevenson's result for $\alpha_R(Q)$.

between the physical cross sections can be written in the forms:

$$\frac{R_{e^+e^-}(s)}{3\sum e_q^2} \frac{\int_0^1 dx g_1^p(x,Q^2) - g_1^n(x,Q^2)}{\frac{1}{3}g_A/g_V} = 1 - \Delta\beta_0 \hat{a}^3 \quad (9)$$

and

$$\frac{R_{e^+e^-}(s)}{3\sum e_q^2} \frac{\int_0^1 dx F_3^{\nu p}(x,Q^2) + F_3^{\bar{\nu}p}(x,Q^2)}{6} = 1 - \Delta\beta_0 \hat{a}^3, \quad (10)$$

provided that the annihilation energy in $R_{e^+e^-}(s)$ and the momentum transfer $Q$ appearing in the deep inelastic structure functions are commensurate at NLO: $\sqrt{s} = Q^* = Q\exp[\frac{7}{4} - 2\zeta_3 + (\frac{11}{96} + \frac{7}{3}\zeta_3 - 2\zeta_3^2 - \frac{\pi^2}{24})\beta_0 \hat{a}(Q)]$. The light-by-light correction to the CSR for the Bjorken sum rule vanishes for three flavors. The term $\Delta\beta_0 \hat{a}^3$ with $\Delta = \ell n\,(Q^{**}/Q^*)$ is the third-order correction arising from the difference between $Q^{**}$ and $Q^*$; in practice this correction is negligible: for a typical value $\hat{a} = \alpha_R(Q)/\pi = 0.14$, $\Delta\beta_0 \hat{a}^3 = 0.007$. Thus at the magic energy $\sqrt{s} = Q^*$, the radiative corrections to the Bjorken and GLLS sum rules almost precisely cancel the radiative corrections to the annihilation cross section. This allows a practical test and extension of the Crewther relation to non-conformal QCD.

As an initial test of Eq. (10), we can compare the CCFR measurement [25] of the Gross-Llewellyn Smith sum rule $1 - \hat{\alpha}_{F_3} = \frac{1}{6}\int_0^1 dx[F_3^{\nu p}(x,Q^2) + F_3^{\bar{\nu}p}(x,Q^2)] = \frac{1}{3}(2.5 \pm 0.13)$ at $Q^2 = 3$ GeV$^2$ and the parameterization of the annihilation data [24] $1 + \hat{\alpha}_R = R_{e^+e^-}(s)/3\sum e_q^2 = 1.20$. at the commensurate scale $\sqrt{s} = Q^* = 0.38\,Q = 0.66$ GeV. The product is $(1+\hat{\alpha}_R)(1-\hat{\alpha}_{F_3}) = 1.00 \pm 0.04$, which is a highly nontrivial check of the theory at very low physical scales. More recently, the E143 [26] experiment at SLAC has reported a new value for the Bjorken sum rule at $Q^2 = 3$GeV$^2$: $\Gamma_1^p - \Gamma_1^n =$

$0.163 \pm 0.010 (\text{stat}) \pm 0.016 (\text{syst})$. The Crewther product in this case is also consistent with QCD: $(1 + \hat{\alpha}_R)(1 - \hat{\alpha}_{g_1}) = 0.93 \pm 0.11$.

Commensurate scale relations such as the generalized Crewther relation discussed here open up additional possibilities for testing QCD. One can compare two observables by checking that their effective charges agree both in normalization and in their scale dependence. The ratio of commensurate scales $\lambda_{A/B}$ is fixed uniquely: it ensures that both observables $A$ and $B$ pass through heavy quark thresholds at precisely the same physical point. The same procedure can be applied to multi-scale problems; in general, the commensurate scales $Q^*, Q^{**}$, etc. will depend on all of the available scales.

An important computational advantage is that one only needs to compute the flavor dependence of the higher order terms in order to specify the lower order scales in the commensurate scale relations. We have shown [1] that in many cases the application of the NLO BLM formulas to relate known physical observables in QCD leads to results with surprising elegance and simplicity. The commensurate scale relations for some of the observables ($\alpha_R, \alpha_\tau, \alpha_{g_1}$ and $\alpha_{F_3}$) are universal in the sense that the coefficients of $\hat{\alpha}_s$ are independent of color; in fact, they are the same as those for Abelian gauge theory. Thus much information on the structure of the non-Abelian commensurate scale relations can be obtained from much simpler Abelian analogs. In fact, in the examples we have discussed here, the non-Abelian nature of gauge theory is reflected in the $\beta$-function coefficients and the choice of second-order scale $Q^{**}$. The commensurate scale relations between observables can apparently be tested at quite low momentum transfers, even where PQCD relationships would be expected to break down. It is possible that some of the higher twist contributions common to the two observables are also correctly represented by the commensurate scale relations. In contrast, expansions of any observable in $\alpha_{\overline{MS}}(Q)$ must break down at low momentum transfer since $\alpha_{\overline{MS}}(Q)$ becomes singular at $Q = \Lambda_{\overline{MS}}$. (For example, in the 't Hooft scheme where the higher order $\beta_n = 0$ for $n = 2, 3, ...$ , $\alpha_{\overline{MS}}(Q)$ has a simple pole at $Q = \Lambda_{\overline{MS}}$.) The commensurate scale relations allow tests of QCD in terms of finite effective charges without explicit reference to singular schemes such as $\overline{MS}$.

The coefficients in a CSR are identical to the coefficients in a conformal theory where renormalons do not appear [4]. It is thus reasonable to expect that the series expansions appearing in the CSR are convergent when one relates finite observables to each other. Thus commensurate scale relations between observables allow tests of perturbative QCD with higher and higher precision as the perturbative expansion grows.

## 3. Electromagnetic and Axial Moments of Relativistic Bound States [6]

The magnetic moments of non-relativistic bound state systems such as atoms are normally computed by summing the moments of its constituents. The situation is much more interesting and complex for composite systems in QCD where relativistic recoil effects must be taken into account. For example, at infinitely small radius $RM \to 0$ and infinitely high excitation energy, the magnetic moment of any spin-$\frac{1}{2}$ system will become equal to the Dirac moment $e/2M$, as can be shown directly from the Drell-Hearn-Gerasimov (DHG) sum rule [27,28]. More remarkably, one can use a generalization [29] of the DHG sum rule to show [30] that the magnetic and quadrupole moments of any composite spin-one system take on the canonical values $\mu = e/M$ and $Q = -e/M^2$ in the limit of zero bound-state radius or infinite excitation energy. Thus in the strong binding limit, the moments of composite particles coincide with the moments of the gauge particles in the tree-graph approximation to the Standard Model. Although the physical structure of spin-one nuclei, spin-one mesons, and the gauge bosons of the Standard Model are highly disparate, there are other underlying universal features. For example, the ratios of their form factors $G_C(Q^2)/G_M(Q^2)$, and $G_C(Q^2)/G_Q(Q^2)$ at large momentum transfer have similar scaling behavior [30] reflecting the underlying gauge and chiral symmetry of the Standard Model at short distances. In this section we shall investigate the quantitative behavior of axial and electromagnetic moments for both strong and weak binding limit, as well as demonstrate the transition between them.

Although the magnetic and quadrupole moments of composite systems are usually regarded as "static" quantities, they actually require the evaluation of the current matrix elements $< p|j_\mu|p + q >$ which are, respectively, linear and quadratic in the momentum transfer $q$. The contribution to the current matrix elements which are generated by the Wigner boost of the state from its rest frame gives a non-additive spin structure for the current interactions of bound systems, and by itself yields the Dirac contribution $\mu = eS/M$ for systems of spin $S$ and the Standard Model quadrupole moment $Q = -e/M^2$ for spin-one states.

The deuteron and triton are non-relativistic bound state systems; nevertheless, one obtains small but nontrivial finite binding corrections to the standard treatment of their magnetic and quadrupole moments [18]. The Wigner boost also leads to the remarkable result that one obtains a non-zero contribution to the quadrupole moment even if the deuteron has no $D-$wave contribution. The same non-additive spin structure is required to reproduce the low energy theorem for Compton scattering on a composite system as well as the DHG sum rule [27] for polarized photoabsorption cross sections [31]. The kinematical boost contribution can be neglected compared to the dynamical contributions from light constituents $\mu = \mathcal{O}(e/m)$ or internal structure $\mu = \mathcal{O}(eR)$ and $Q = \mathcal{O}(eR^2)$ if $M/m \gg 1$ and $MR \gg 1$. Thus the usual formulas for computing moments from the sum of constituent moments is only strictly valid in the cases of systems such as atoms where the electron mass is small compared to the atomic mass and the Bohr size $R$ is large compared to the Compton scale $1/M$ of the atom.

The light-cone ("front-form") formalism [32] provides a convenient covariant frame-work for evaluating current matrix elements of composite systems [28]. The formalism is independent of the choice of momentum $p^\mu$, and form factors can be calculated from diagonal matrix elements; i.e, the convolution of light-cone wavefunctions with the same particle number $n$. In contrast, in equal-time theory, one needs to consider frame-dependent non-diagonal pair creation matrix elements as well as vacuum creation contributions to the current which are unconstrained by the Fock wavefunctions. The Bethe-Salpeter formalism is covariant, but one needs to evaluate the matrix elements of an infinite number of irreducible kernels, even in the case when one constituent is infinitely heavy.

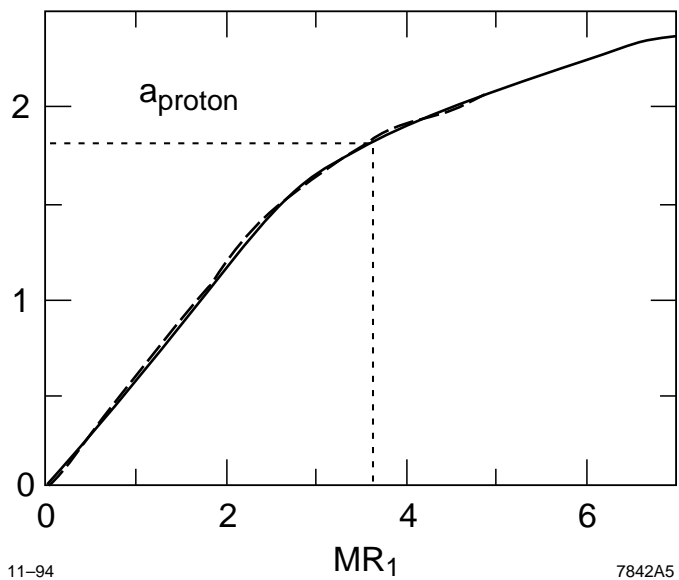

Figure 3: The anomalous magnetic moment $a = F_2(0)$ of the proton as a function of $M_p R_1$: broken line, pole type wavefunction; continuous line, gaussian wavefunction. The experimental value is given by the dotted lines. The prediction of the model is independent of the wavefunction for $Q^2 = 0$.

A three-quark light-cone model can be used to illustrate the functional relationship between the anomalous moment of a proton $a_p$ and its Dirac radius [6]. The value of $R_1^2 = -6 dF_1(Q^2)/dQ^2|_{Q^2=0}$ is varied by changing the size parameters. Figure 3 shows that when one plots the dimensionless observable $a_p$ against the dimensionless observable $MR_1$, the prediction is essentially independent of the assumed power-law or Gaussian form of the three-quark light-cone wavefunction. The only parameter controlling the relation between the dimensionless observables in the light-cone three-quark model is $m/M_p$ which is set to 0.28. For the physical proton radius $M_p R_1 = 3.63$, one obtains the empirical value for $a_p = 1.79$ (indicated by the dotted lines in Fig. 3). The same three-quark model also gives $g_A = 1.25$ for the non-singlet axial coupling in agreement with experiment. The singlet helicity sum $\Delta\Sigma$ for the three quark model is predicted to be $\cong 0.75$. This will be substantially reduced when gluon and sea quark Fock state contributions are included. The relativistic effects also reduce the anomalous magnetic moment and axial coupling by a factor of $\simeq 0.75$ [6,33]. The fact that both the axial coupling and lowest moment of the $g_1$ structure function of a composite system are modified by the Melosh transformation was first pointed out by Bucella, et al. [34], Le Youanc, et al. [35], and Close [33]. An important consistency check of any bound state formalism is the demonstration that the electromagnetic moments of a composite system reproduces the canonical Standard Model values in the point-like limit $MR \to 0$. The light-cone analysis correctly reproduces the correct ultra-relativistic limit for the electromagnetic moments. Thus in the pointlike limit where the threshold for particle excitation $\nu_{th} \to \infty$, even a system as complex as the deuteron acquires the same electromagnetic moments $(Q_d^a \to 0, a_d \to 0)$ as that of the $W$ in the Standard Model.

The light-cone model also predicts that the quark helicity sum $\Delta\Sigma = \Delta u + \Delta d$ and $g_A = \Delta u - \Delta d$ vanishes as a function of the proton radius $R_1$ in a similar way as the anomalous moment vanishes. Since the helicity sum $\Delta\Sigma$ depends on the proton size, it clearly cannot be identified as the vector sum of the rest-frame constituent spins. Actually, $\Delta q$ refers to the difference of helicities at fixed light-cone time or at infinite momentum; it cannot be identified with $q(s_z = +\frac{1}{2}) - q(s_z = -\frac{1}{2})$, the spin carried by each quark flavor in the proton rest frame in the equal time formalism [36,6]. In fact, $\Delta q$ vanishes as $R_1 \to 0$. Similar results are obtained for spin-one systems: At small deuteron radius the light-cone model predicts not only a vanishing anomalous moment but also $\lim_{R \to 0} g_A(M_d R) = 0$. As shown by Ma and Zhang [36] the Melosh rotation generated by the internal transverse momentum spoils the usual identification of the $\gamma^+ \gamma_5$ quark current matrix element with the total rest-frame spin projection $s_z$, thus resulting in a reduction of $g_A$. One can understand this physically: in the zero radius limit the internal transverse momenta become infinite and the nucleon helicities become completely disoriented.

These results have important implications for theories in which leptons, quarks, or gauge particles are composite at short distances. If the internal scale of such a theory is sufficiently high, then the DHG sum rule [27] guarantees that the magnetic and quadrupole couplings of the composite states are indistinguishable from those of the Standard Model. In addition, one finds in the light-cone models that the axial couplings of composite spin-one systems vanish in the point-like limit. In the Standard Model the parity-violating Gamow-Teller axial couplings of the $W$ and $Z$ vanish at tree level. Thus, even though composite spin-one systems are not gauge fields, their couplings can simulate the canonical axial and electromagnetic moments of the Standard Model provided they are sufficiently compact. This is interesting from the phenomenological point of view, since it keeps open the possibility that the $Z$ and $W$ vector bosons of the Standard Model could be composite provided their internal scale is sufficiently small and their excitation energies are sufficiently high [37]. On the other hand, the light-cone Fock state description predicts $g_A \to 0$ for composite spin-$\frac{1}{2}$ systems in the point-like limit, whereas the canonical axial coupling in the Standard Model is $g_A = 1$ for elementary spin-$\frac{1}{2}$ fields. It thus remains an open question whether a consistent dynamical model of composite leptons and quarks [38] can be formulated which can simultaneously simulate their observed magnetic moment and axial couplings.

## 4. Precision limits on Anomalous Couplings of the $W$ and $Z$ [9]

The Dirac value $g = 2$ for the magnetic moment $\mu = geS/2M$ of a particle of charge $e$, mass $M$, and spin $S$, plays a special role in quantum field theory. As shown by Weinberg [39] and Ferrara $et.$ $al$ [40], the canonical value $g = 2$ gives an effective Lagrangian which has maximally convergent high energy behavior for fields of any spin. In the case of the Standard Model, the anomalous magnetic moments $\mu_a = (g-2)eS/2M$ and anomalous quadrupole moments $Q_a = Q + e/M^2$ of the fundamental fields vanish at tree level, ensuring a quantum field theory which is perturbatively renormalizable. However, as discussed in the previous section, one can use the DHG sum rule [27] to show that the magnetic and quadrupole moments of spin-$\frac{1}{2}$ or spin-1 bound states approach the canonical values $\mu = eS/M$ and $Q = -e/M^2$ in the zero radius limit $MR \to 0$ [28,6,30], independent of the internal dynamics. Deviations from the predicted values will thus reflect new physics and interactions such as virtual corrections from supersymmetry or an underlying composite structure.

The canonical values $g = 2$ and $Q = -e/M^2$ lead to a number of important phenomenological consequences: (1) The magnetic moment of a particle with $g = 2$ processes with the same frequency as the Larmor frequency in a constant magnetic field. This synchronicity is a consequence of the fact that the electromagnetic spin currents can be formally generated by an infinitesimal Lorentz transformation [41,42]. (2) The forward helicity-flip Compton amplitude for a target with $g = 2$ vanishes at zero energy [43]. (3) The Born amplitude for a photon radiated in the scattering of any number of incoming and outgoing particles with charge $e_i$ and four-momentum $p_i^\mu$ vanishes at the kinematic angle where all the ratios $e_i/p_i \cdot k$ are simultaneously equal [42]. For example, the Born cross section $d\sigma/\cos\theta_{cm}(u\overline{d} \to W^+\gamma)$ vanishes identically at an angle determined from the ratio of charges: $\cos\theta_{cm} = e_d/e_{W+} = -1/3$ [44]. Such "radiative amplitude zeroes" or "null zones" occur at lowest order in the Standard Model because the electromagnetic spin currents of the quarks and the vector gauge bosons are all canonical.

The vanishing of the forward helicity-flip Compton amplitude at zero energy for the canonical couplings, together with the optical theorem and dispersion theory, leads to a superconvergent sum rule; *i.e.*, a zero value for the DHG sum rule. This remarkable observation was first made for quantum electrodynamics and the elec-troweak theory by Altarelli, Cabibbo and Maiani [7]. More generally, one can use a quantum loop [8] expansion to show that the logarithmic integral of the spin-dependent part of the photoabsorption cross section

$$\int_{\nu_{th}}^{\infty} \frac{d\nu}{\nu} \Delta\sigma_{\text{Born}}(\nu) = 0 \tag{11}$$

for any $2 \to 2$ Standard Model process $\gamma a \to bc$ in the classical, tree graph approximation. The particles $a, b, c$ and $d$ can be leptons, photons, gluons, quarks, elementary Higgs particles, supersymmetric particles, etc. We also can extend the sum rule to certain virtual photon processes. Here $\nu = p \cdot q/M$ is the laboratory energy and $\Delta\sigma(\nu) = \sigma_P(\nu) - \sigma_A(\nu)$ is the difference between the photoabsorption cross section for parallel and antiparallel photon and target helicities. The sum rule receives nonzero contributions in higher order perturbation theory in the Standard Model from both quantum loop corrections and higher particle number final states. Similar arguments also imply that the DHG integral vanishes for virtual photoabsorption processes such as $\ell\gamma \to \ell Q\overline{Q}$ and $\ell g \to \ell Q\overline{Q}$, the lowest order sea-quark contribution to polarized deep inelastic photon and hadron structure functions. Note that the integral extends to $\nu = \nu_{th}$, which is generally beyond the usual leading twist domain.

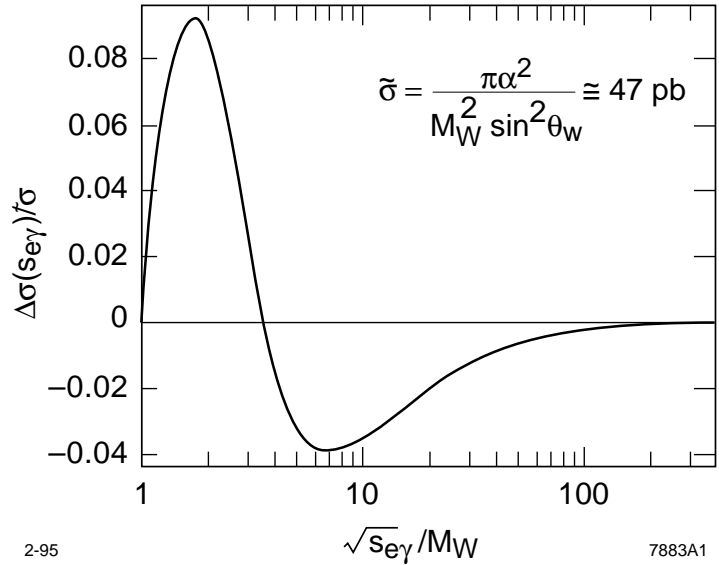

Figure 4: The Born cross section difference $\Delta\sigma$ for the Standard Model process $\gamma e \to W\nu$ for parallel minus antiparallel electron/photon helicities as a function of $\log \sqrt{s}_{e\gamma}/M_W$ The logarithmic integral of $\Delta\sigma$ vanishes in the classical limit.

We can use Eq. (11) as a new way to test the canonical couplings of the Standard Model and to isolate the higher order radiative corrections. The sum rule also provides a non-trivial consistency check on calculations of the polarized cross sections.

Probably the most interesting application and test of the Standard Model is to the reactions $\gamma\gamma \to q\bar{q}$, $\gamma e \to W\nu$ and $\gamma e \to Ze$ which can be studied in high energy polarized electron-positron colliders with back-scattered laser beams. In contrast to the timelike process $e^+e^- \to W^+W^-$, the $\gamma\gamma$ and $\gamma e$ reactions are sensitive to the anomalous moments of the gauge bosons at $q^2 = 0$. The cancellation of the positive and negative contributions [45] of $\Delta\sigma(\gamma e \to W\nu)$ to the DHG integral is evident in Fig. 4.

We can also exploit the fact that the vanishing of the logarithmic integral of $\Delta\sigma$ in the Born approximation also implies that there must be a center-of-mass energy, $\sqrt{s}_0$, where the polarization asymmetry $A = \Delta\sigma/\sigma$ possesses a zero, i.e., where $\Delta\sigma(\gamma e \to W\nu)$ reverses sign [9]. We find strong sensitivity of the position of this zero or "crossing point" (which occurs at $\sqrt{s}_{\gamma e} = 3.1583\ldots M_W \simeq 254$ GeV in the SM) to modifications of the SM trilinear $\gamma WW$ coupling. Given reasonable assumptions for the luminosity and energy range for the Next Linear Collider(NLC), the zero point, $\sqrt{s}_0$, of the polarization asymmetry may be determined with sufficient precision to constrain the anomalous couplings of the $W$ to better than the 1% level at 95% CL. Since the zero occurs at rather modest energies where the unpolarized cross section is near its maximum, an electron-positron collider with $\sqrt{s} = 320 - 400$ GeV is sufficient, whereas other techniques aimed at probing the anomalous couplings through the $\gamma e \to W\nu$ process require significantly larger energies. In addition to the fact that only a limited range of energy is required, the polarization asymmetry measurements have the advantage that many of the systematic errors cancel in taking cross section ratios. This technique can clearly be generalized to other higher order tree-graph processes in the Standard Model and supersymmetric gauge theory. The position of the zero in the photoabsorption asymmetry thus provides an additional weapon in the arsenal used to probe anomalous trilinear gauge couplings.

## 5. Acknowledgments

The work on the generalized Crewther relation reported in Section 2 here has greatly benefitted from discussions with A. Kataev and G. Gabadadze. The results reported in Section 3 are based on collaborations with F. Schlumpf. Section 4 is based on collaborations with I. Schmidt and T. Rizzo. We also thank V. Braun, M. Gill, J. Hiller, P. Huet, C.-R. Ji, G. P. Lepage, G. Mirabelli, A. Mueller, D. Müller, Alex Pang, O. Puzyrko, and W.-K. Wong, for helpful discussions. This work is supported in part by the Department of Energy, contract DE–AC03–76SF00515 and contract DE–FG03–93ER–40792.